\documentclass{aa}  
\usepackage{xcolor}
\usepackage{graphicx}
\usepackage{txfonts}
\begin{document}

   \title{Timing Argument take on the Milky Way and Andromeda past-encounter}

   \author{David ~Benisty\inst{1,2,3}}

   \institute{Frankfurt Institute for Advanced Studies, Ruth-Moufang-Strasse~1, 60438 Frankfurt am Main, Germany
 \\
 \and DAMTP, Centre for Mathematical Sciences, University of Cambridge, Wilberforce Road, Cambridge CB3 0WA, UK
 \\ \and Institute of Astronomy, University of Cambridge, Madingley Road, Cambridge, CB3 0HA, UK}

  \abstract
 {The two-body problem of $M31$ and the Milky Way (MW) galaxies with a Cosmological Constant background is studied, with emphasis on the possibility that they experienced Past Encounter (PE). PE are possible only for non-zero transverse velocity and their viability is subject to observations of the imprints of such near collisions.  By implementing the Timing Argument (TA) for two isolated point bodies, it is shown that if $M{31}$ and the MW have had PE, then the predicted mass of the Local Group (LG) should be twice larger. Since the predicted mass is too large, the MW and $M31$ galaxies should have collided at $\sim 8 Gys$. Therefore, the TA analysis show that PE is not possible for the Local Group (LG) system. }

   \keywords{Local Group Mass - Cosmological Constant - Past Encounters}

   \maketitle

\section{Introduction}
Knowledge of the mass of the Local Group is a crucially important. However, its observational determination is challenging as a consequence of the fact that the mass of the LG is dominated by dark matter which cannot be directly observed. The Timing Argument (TA) was proposed in \cite{1989ApJ345108P,1981Obs101111L} for the Local Group, as one of the historical probes of the missing mass problem. The TA assumes that Milky Way ($MW$) and Andromeda ($M31$) have been approaching each other despite cosmic expansion. In its simplest version, the LG consists of the $MW$ and $M31$ as two isolated point masses. Briefly after the big bang these two galaxies must have been in the same place with zero separation. Due to the Hubble expansion these two galaxies moved apart. After couple of billion years, they slowed down and then moved towards each other again, as a consequence of the gravitational pull.

Earlier estimations for the LG mass have been done with different methods:
considering simulations
\cite{Carlesi:2016eud,Gonzalez:2013pqa,Penarrubia:2014oda,Li:2007eg,2018MNRAS.477.4768B,Hartl:2021aio},
the TA as much as the viral theorem
\cite{vanderMarel:2012xp,Chernin:2009ms}, numerical action method
\cite{Phelps:2013rra}, machine learning \cite{McLeod:2016bjk} and Likelihood-Free Inference \cite{Lemos:2020vhj}. The
estimations predict that the mass of the LG should be around $10^{12}$ solar masses ($M_{\odot}$). \cite{Zhao:2013uya,Benisty:2020kys} test the LG mass in Modified Newtonian Dynamics gravity. This paper investigates the effects of PE on the predicted mass of the LG.

The structure of this paper is as follows: Section \ref{sec:form} discusses the formalism of two-body motion with a Cosmological Constant. Section \ref{sec:Meth} explains the numerical method. Section \ref{sec:AM} explains the analysis considering possible Past Encounter (PE). Section \ref{sec:Dis} discusses the results.

\begin{figure}[t!]
 	\centering
\includegraphics[width=0.5\textwidth]{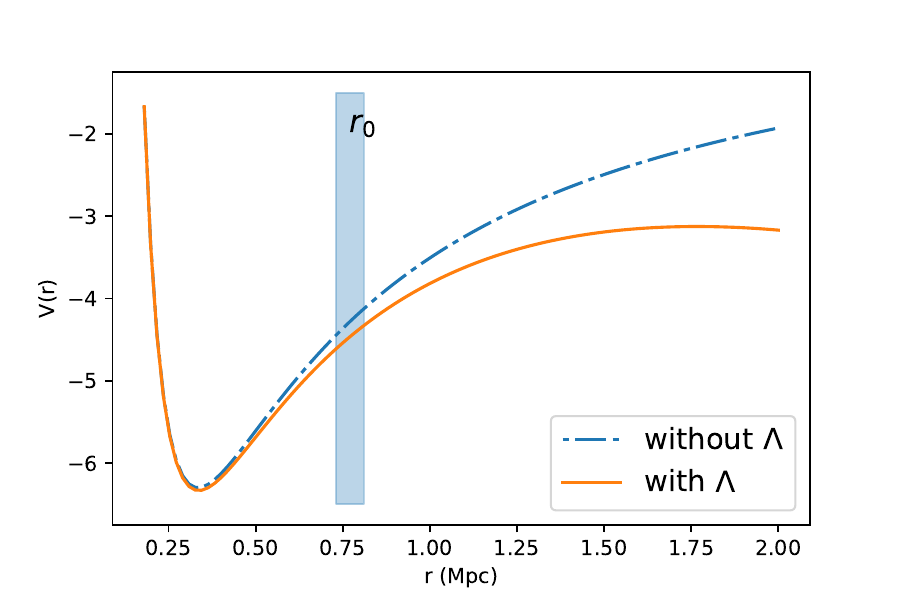}
\caption{\it{The effective potential vs. radius for two-body system with or without a Cosmological Constant. The distance of M31 towards us is noted in the plot with a blue range, with $1 \sigma$ range.}}
 	\label{fig:1}
  \end{figure}

\section{Two Body Problem}
\label{sec:form}
The Cosmological Constant domination is considered to govern at cosmological scales but also in the LG scale \cite{Chernin:2000pq,Baryshev:2000kw,Chernin:2001nu,Karachentsev:2003eh,Chernin:2003qd,Teerikorpi:2005zh,Chernin:2009ms,Chernin:2006dy,Teerikorpi:2010zz,Chernin:2015nna,Silbergleit:2019oyx,Benisty:2020kys}. Hence we include the Cosmological Constant contribution in our analysis. The center of mass coordinate system is defined by the relative distance $r$ and the relative velocity $v$. The masses are replaced by the total mass $M$. In polar coordinate system $(r,\varphi)$ the relative distance variation reads \cite{Emelyanov:2015ina,Carrera:2006im}:
\begin{equation}\label{ENL}
\ddot{r} = \frac{L^2}{r^3}-\frac{GM}{r^2} +  \frac{1}{3}\Lambda c^2 \, r,
\end{equation}
where $L$ is the conserved angular momentum per mass ($\vec{L} = \vec{r} \times \vec{v}$), or:
\begin{equation}\label{am}
L=r^2 \dot\varphi = r \, v_{\textit{tan}},
\end{equation}
where $v_{\textit{tan}}$ is the tangential velocity. Since the potential is time independent, the associated conserved energy reads:
\begin{equation}
E = \frac{1}{2} \dot{r}^2 + V(r),
\end{equation}
with the effective potential:
\begin{equation}
V\left(r\right) = \frac{L^2}{2r^2} - \frac{GM}{r} - \frac{\Lambda c^2}{12} r^2.
\end{equation}

\begin{figure}[t!]
 	\centering
\includegraphics[width=0.44\textwidth]{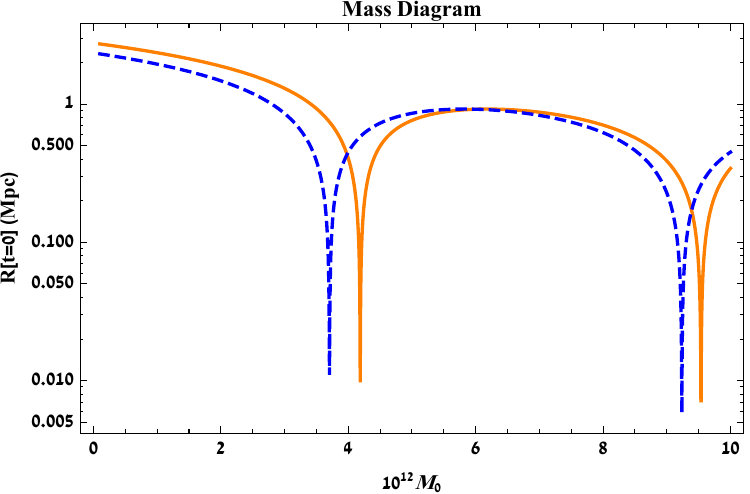}
\caption{\it{The evaluated distance between Milky Way and $M\text{31}$ at $t=0$ for different masses of the LG. The model predicts the correct mass when $r(t=0) = 0$. The first minimum ($4-5\, \cdot 10^{12} M_{\odot}$) points the predicted mass according to the TA. The second minimum points the predicted mass with one past encounter ($9-10\, \cdot 10^{12} M_{\odot}$). The blue dashed line corresponds to GR without $\Lambda$ and the orange smooth corresponds to GR with $\Lambda$.}}
 	\label{fig:2}
  \end{figure}
Fig. \ref{fig:1} presents the potential $V(r)$. There are two extreme points: The closer one is a minimum and the other is a maximum. For large distances the repulsive force arising from the Cosmological Constant makes impossible any bound structure so the Cosmological Constant gives us a bound on the maximum size of the bound systems.

\section{The Method}
\label{sec:Meth}
We restrict ourselves to a simple basis of the TA as an isolated, two dimensional system. Galaxies are modeled as point masses. Moreover, as the galaxy pairs are isolated, there are no external gravitational fields. The "initial condition" of the model is $r(t=0) = 0$, which corresponds to the "big bang".  In order to calculate the mass of the Local group we change the direction of time by considering the opposite direction of the measured velocity. We evaluate the measured distance of $M31$ to obtain what should be the distance at the "big bang" for different LG masses. Figure (\ref{fig:2}) presents the distance for the big bang for different masses of LG. When the curve approaches minimum ($r \rightarrow 0$) yielding the predicted mass. The minimum recognizes the mass of the LG as the model predicts. The location and the velocity of the M31 are:
\begin{equation*}
\begin{split}
r_{m_{31}} = 0.77 \pm 0.04 \, {\text{Mpc}}, \\ v_{rad} = -109.3 \pm 4.4 \, {km/sec},  \\
v_{tan}(t_u) = 17 \, {km/sec},
\end{split}
\end{equation*}
with a $1\,\sigma$ confidence, with the region of $V_{tan} \leq 34.3 km/sec $ which measured in \cite{vanderMarel:2012xp}. $v_{rad}$ is the radial velocity towards the Milky Way galaxy. The cosmological constant is being:
\begin{equation*}
\Lambda = (4.24 \pm 0.11) \cdot 10^{-66} \, eV^2
\end{equation*}
determined by the latest Planck measurements \cite{Aghanim:2018eyx}. When we estimate the initial distance at the big bang, we use the age of the universe determined also by the latest Planck measurements.

The acceleration terms Eq. (\ref{ENL}) are in the same order if we set the mass to be around $M \sim 10^{12} M_{\odot}$ and and the distance between the galaxies around $r \sim 1 Mpc$:
\begin{equation}
- \frac{G M}{r^2} \approx 6 \cdot 10^{-9} \frac{m}{sec^2}, \quad \frac{1}{3}\Lambda c^2 r^2 = 10^{-9} \frac{km}{sec^2}.
\end{equation}
Both terms are in the same order. Therefore, one has take into account also the expansion effect even thought the scale is very short.

\cite{Eingorn:2012dg,Eingorn:2012jm,Partridge:2013dsa,Gonzalez:2013pqa,McLeod:2016bjk,McLeod:2019cfg} originally show that the Cosmological Constant effect in the LG scales ($1 Mpc$) changes the predicted mass of LG to be larger. The mass of the LG without the Cosmological Constant is $M_{LG} = \left( 4.17 \pm 0.89 \right) \cdot 10^{12} \, M_{\odot}$, and with the Cosmological Constant presence we get: $M_{LG} = \left( 4.73 \pm 1.03 \right) \cdot 10^{12}  \, M_{\odot}$. The difference is not negligible. 

\begin{figure*}[t!]
 	\centering
\includegraphics[width=1.\textwidth]{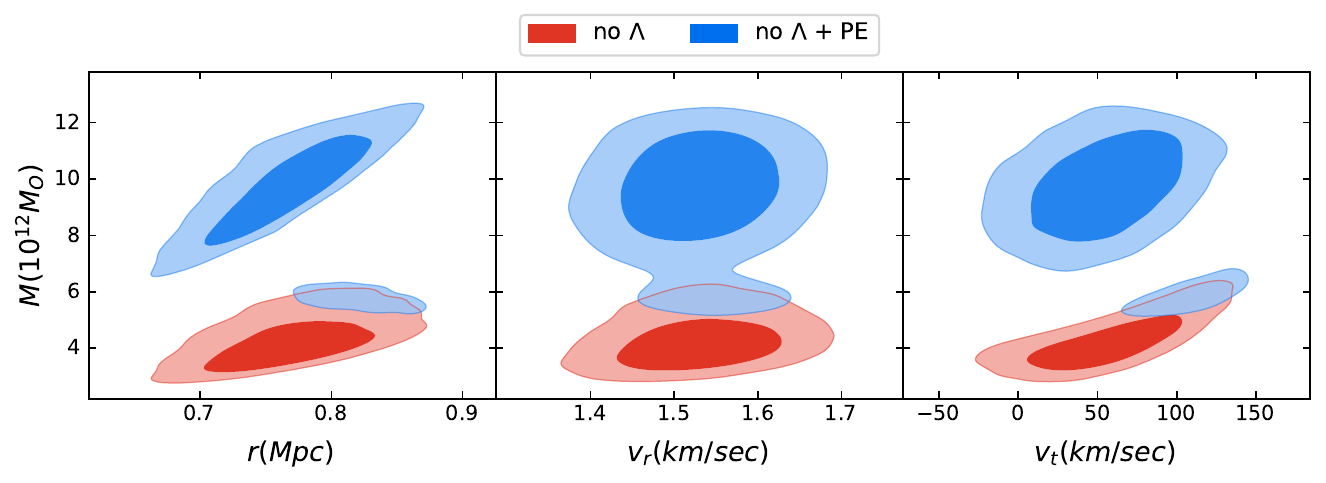}
\\
\includegraphics[width=1\textwidth]{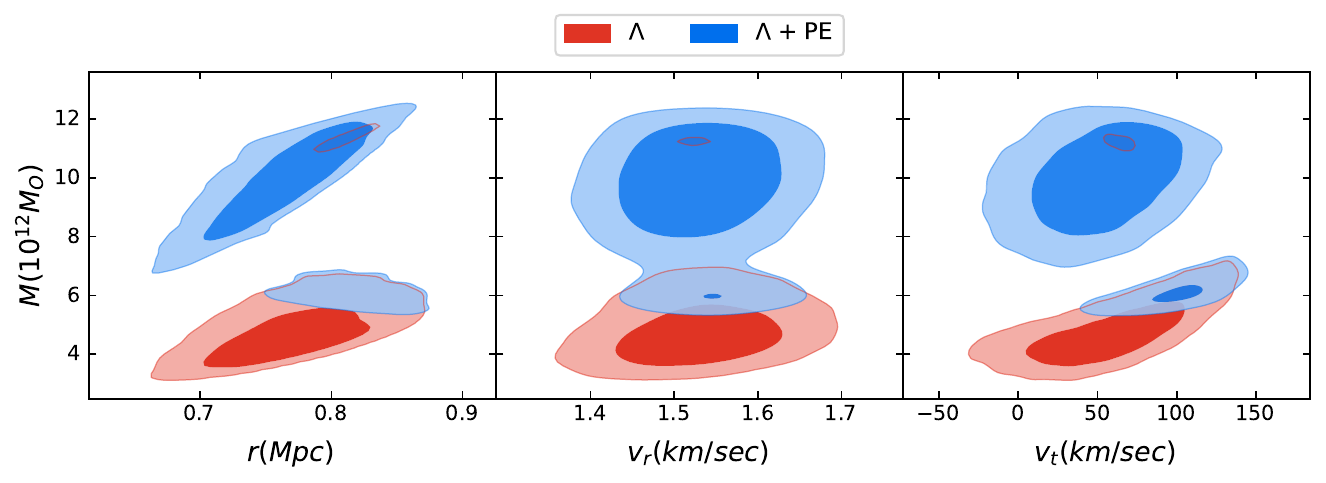}
\caption{\it{The corner plots for the predicted mass of the LG vs. the measured distance towards M31 and the measured velocities. The upper panel shows the results without $\Lambda$ and the lower panel shows the results with $\Lambda$. }}
 	\label{fig:post}
  \end{figure*}

\section{Results}
\label{sec:AM}
The presence of the tangential velocity yields the possibility for PE. The masses diagram vs. the evaluated distance in the big bang ($t = 0$) is presented in Fig. (\ref{fig:2}). We see that there are several masses that fit to the initial condition ($r(t) \rightarrow 0$). The left minimum is the model perdition without any PE. The right minimum is the model perdition with one PE. 

\begin{table}[t!]
 	\centering
\begin{tabular}{| c | c |}
\hline\hline
Case & $M (10^{12} M_\odot)$  \\
\hline\hline 
no $\Lambda$, no PE  & $4.13 \pm 0.78$
\\
\hline
with $\Lambda$, no PE & $4.61 \pm 1.39$
\\
\hline
no $\Lambda$, with PE & $9.63 \pm 2.14$
\\
\hline
with $\Lambda$, with PE & $9.77 \pm 2.47$
\\
\hline\hline  
\end{tabular} 
\caption{\it{Summary of predicted mass of the LG for with or without $\Lambda$ and with or without PE.  }}
\centering
\label{tabel:sum}
\end{table}

Table \ref{tabel:sum} summarizes the results for the predicted mass of LG with or without PE. Fig \ref{fig:post} shows the corresponding cornerplots. With no Cosmological Constant background the mass is $4.13 \pm 0.78\, M_{\odot}$. With a Cosmological Constant background the mass is $4.61 \pm 1.39\, M_{\odot}$. For any additional encounter, the predicted mass is larger. Without $\Lambda$ the predicted mass is $9.63 \pm 2.14 \,  M_{\odot}$ and with $\Lambda$ is $9.77 \pm 2.47  \,  M_{\odot}$.

\begin{figure}
 	\centering
\includegraphics[width=0.44\textwidth]{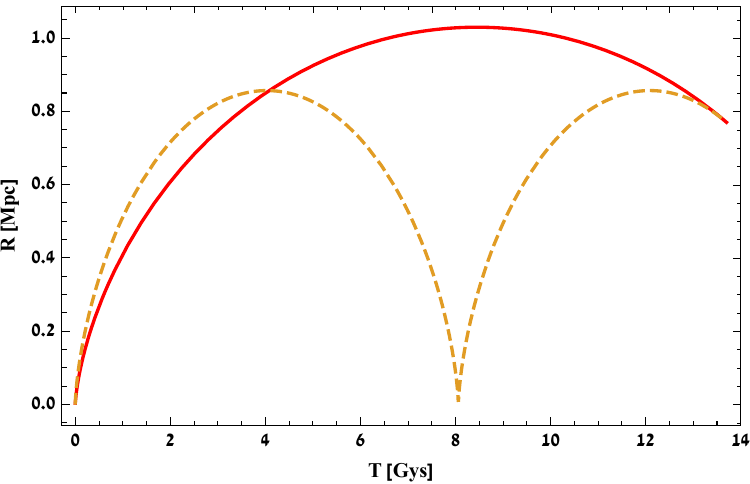}
\caption{\it{The distance between the galaxies vs. time without or with with PE. The The minimal point around $\sim 8.1 Gys$ shows the galaxies fly by and  }}
 	\label{fig:motion}
  \end{figure}

In order to track the actual motion of the galaxies we evaluate a numerical solution for the distance between the galaxies with the extracted mass. Fig \ref{fig:motion} shows the numerical solution of the distance vs. time, with or without PE. The smooth line shows the solution from the big bang point $r(t = 0) = 0$ to the final stage $r (t = 13.7 Gys) = 0.77 Mpc$. The dashed line shows the numerical solution for the larger mass and yields PE. Around $\sim 8.1$ Gys after the big bang the galaxies has had to fly by each other in the case of PE.

The model assumes that the galaxies are point masses. This assumption is correct since the size of the galaxies is negligible partially to the distance between the galaxies at most of the history of the universe, where there is no PE (see fig \ref{fig:motion}). However, in the case of the PE, the masses are predicted to be twice larger from the TA and also has had to fly by each other. In the time the galaxies fly by, the strong Dynamical Friction (DF) from the dark matter cause the galaxies to merge \cite{1987gady,Hammer:2007ki}. \cite{Cox:2007nt,Conselice} use N-body simulation to track the evolution of the LG. Also simulation shows that DF between the galaxies would lead to the eventual merger of the LG. Therefore from the simple TA approximation, it is possible to rules out the existence of PE as the other simulation suggests.

\section{Summery}
\label{sec:Dis}
\begin{figure}[t!]
 	\centering
\includegraphics[width=0.46\textwidth]{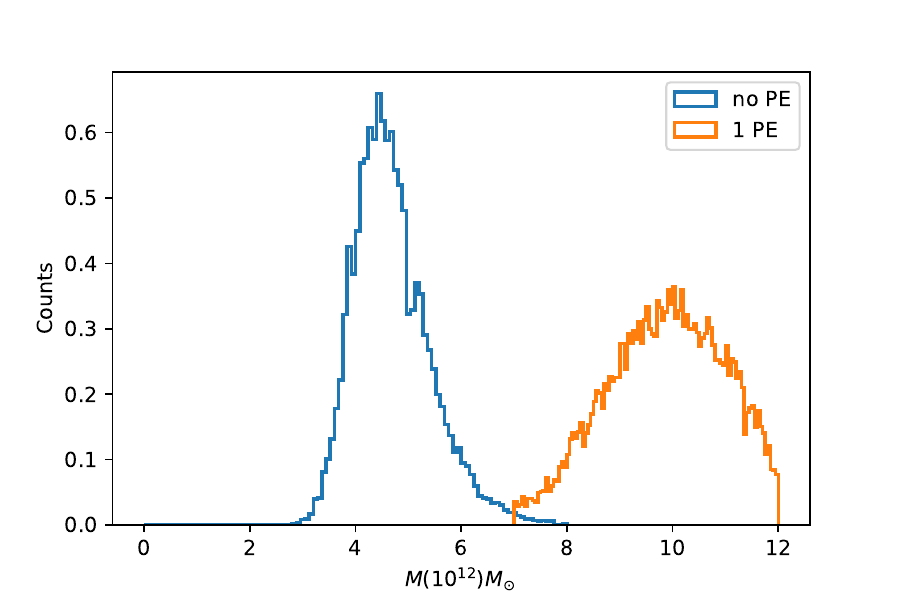}
\caption{\it{Final comparison between the mass of the LG without and with one PE.}}
 	\label{fig:fin}
  \end{figure}

In this paper we relate the mass of LG with the a possible PE using the TA. we are treating the MW and M31 galaxies as point particles that emerge briefly after the big bang at a very small distance. In order to have the same final conditions for the M31 distance and velocity at the present time one can extract the mass for the LG. Fig \ref{fig:fin} shows the final comparison between the masses with or without PE.

M31 and Milky Way are considered as interacting point-like gravitating systems. This is partially correct because it depends on how the galaxy size is negligible with respect to the overall size of the LG. Taking into account the possibility for PE yielding twice larger mass. Since the galaxies in earlier times has had to fly by, the galaxies should be merged because of the larger mass with dynamical friction presence. Therefore, from a simple analysis of the TA the possibly for a PE is ruled out.

\section*{Acknowledgments}
I thank to Ofer Lahav, Pablo Lemos, Yehuda Hofman, Michael McLeod, Mordehai Milgrom, Indranil Banik and Hongsheng Zhao for stimulating discussions. This article is supported by COST Action CA15117 "Cosmology and Astrophysics Network for Theoretical Advances and Training Action" (CANTATA) of the COST (European Cooperation in Science and Technology) and the COST actions CA15117 and CA-16104.

\bibliography{ref}

\begin{thebibliography}{36}%
\makeatletter
\providecommand \@ifxundefined [1]{%
 \@ifx{#1\undefined}
}%
\providecommand \@ifnum [1]{%
 \ifnum #1\expandafter \@firstoftwo
 \else \expandafter \@secondoftwo
 \fi
}%
\providecommand \@ifx [1]{%
 \ifx #1\expandafter \@firstoftwo
 \else \expandafter \@secondoftwo
 \fi
}%
\providecommand \natexlab [1]{#1}%
\providecommand \enquote  [1]{``#1''}%
\providecommand \bibnamefont  [1]{#1}%
\providecommand \bibfnamefont [1]{#1}%
\providecommand \citenamefont [1]{#1}%
\providecommand \href@noop [0]{\@secondoftwo}%
\providecommand \href [0]{\begingroup \@sanitize@url \@href}%
\providecommand \@href[1]{\@@startlink{#1}\@@href}%
\providecommand \@@href[1]{\endgroup#1\@@endlink}%
\providecommand \@sanitize@url [0]{\catcode `\\12\catcode `\$12\catcode
  `\&12\catcode `\#12\catcode `\^12\catcode `\_12\catcode `\%12\relax}%
\providecommand \@@startlink[1]{}%
\providecommand \@@endlink[0]{}%
\providecommand \url  [0]{\begingroup\@sanitize@url \@url }%
\providecommand \@url [1]{\endgroup\@href {#1}{\urlprefix }}%
\providecommand \urlprefix  [0]{URL }%
\providecommand \Eprint [0]{\href }%
\providecommand \doibase [0]{http://dx.doi.org/}%
\providecommand \selectlanguage [0]{\@gobble}%
\providecommand \bibinfo  [0]{\@secondoftwo}%
\providecommand \bibfield  [0]{\@secondoftwo}%
\providecommand \translation [1]{[#1]}%
\providecommand \BibitemOpen [0]{}%
\providecommand \bibitemStop [0]{}%
\providecommand \bibitemNoStop [0]{.\EOS\space}%
\providecommand \EOS [0]{\spacefactor3000\relax}%
\providecommand \BibitemShut  [1]{\csname bibitem#1\endcsname}%
\let\auto@bib@innerbib\@empty
\bibitem [{\citenamefont {{Peebles}}\ \emph {et~al.}(1989)\citenamefont
  {{Peebles}}, \citenamefont {{Melott}}, \citenamefont {{Holmes}},\ and\
  \citenamefont {{Jiang}}}]{1989ApJ345108P}%
  \BibitemOpen
  \bibfield  {author} {\bibinfo {author} {\bibfnamefont {P.~J.~E.}\
  \bibnamefont {{Peebles}}}, \bibinfo {author} {\bibfnamefont {A.~L.}\
  \bibnamefont {{Melott}}}, \bibinfo {author} {\bibfnamefont {M.~R.}\
  \bibnamefont {{Holmes}}}, \ and\ \bibinfo {author} {\bibfnamefont {L.~R.}\
  \bibnamefont {{Jiang}}},\ }\href {\doibase 10.1086/167885} {\bibfield
  {journal} {\bibinfo  {journal} {Astrophys. J.}\ }\textbf {\bibinfo {volume}
  {345}},\ \bibinfo {pages} {108} (\bibinfo {year} {1989})}\BibitemShut
  {NoStop}%
\bibitem [{\citenamefont {{Lynden-Bell}}(1981)}]{1981Obs101111L}%
  \BibitemOpen
  \bibfield  {author} {\bibinfo {author} {\bibfnamefont {D.}~\bibnamefont
  {{Lynden-Bell}}},\ }\href@noop {} {\bibfield  {journal} {\bibinfo  {journal}
  {The Observatory}\ }\textbf {\bibinfo {volume} {101}},\ \bibinfo {pages}
  {111} (\bibinfo {year} {1981})}\BibitemShut {NoStop}%
\bibitem [{\citenamefont {Carlesi}\ \emph {et~al.}(2017)\citenamefont
  {Carlesi}, \citenamefont {Hoffman}, \citenamefont {Sorce},\ and\
  \citenamefont {Gottlöber}}]{Carlesi:2016eud}%
  \BibitemOpen
  \bibfield  {author} {\bibinfo {author} {\bibfnamefont {E.}~\bibnamefont
  {Carlesi}}, \bibinfo {author} {\bibfnamefont {Y.}~\bibnamefont {Hoffman}},
  \bibinfo {author} {\bibfnamefont {J.~G.}\ \bibnamefont {Sorce}}, \ and\
  \bibinfo {author} {\bibfnamefont {S.}~\bibnamefont {Gottlöber}},\ }\href
  {\doibase 10.1093/mnras/stw3073} {\bibfield  {journal} {\bibinfo  {journal}
  {Mon. Not. Roy. Astron. Soc.}\ }\textbf {\bibinfo {volume} {465}},\ \bibinfo
  {pages} {4886} (\bibinfo {year} {2017})},\ \Eprint
  {http://arxiv.org/abs/1611.08078} {arXiv:1611.08078 [astro-ph.GA]}
  \BibitemShut {NoStop}%
\bibitem [{\citenamefont {Gonzalez}\ \emph {et~al.}(2014)\citenamefont
  {Gonzalez}, \citenamefont {Kravtsov},\ and\ \citenamefont
  {Gnedin}}]{Gonzalez:2013pqa}%
  \BibitemOpen
  \bibfield  {author} {\bibinfo {author} {\bibfnamefont {R.~E.}\ \bibnamefont
  {Gonzalez}}, \bibinfo {author} {\bibfnamefont {A.~V.}\ \bibnamefont
  {Kravtsov}}, \ and\ \bibinfo {author} {\bibfnamefont {N.~Y.}\ \bibnamefont
  {Gnedin}},\ }\href {\doibase 10.1088/0004-637X/793/2/91} {\bibfield
  {journal} {\bibinfo  {journal} {Astrophys. J.}\ }\textbf {\bibinfo {volume}
  {793}},\ \bibinfo {pages} {91} (\bibinfo {year} {2014})},\ \Eprint
  {http://arxiv.org/abs/1312.2587} {arXiv:1312.2587 [astro-ph.CO]} \BibitemShut
  {NoStop}%
\bibitem [{\citenamefont {Peñarrubia}\ \emph {et~al.}(2014)\citenamefont
  {Peñarrubia}, \citenamefont {Ma}, \citenamefont {Walker},\ and\
  \citenamefont {McConnachie}}]{Penarrubia:2014oda}%
  \BibitemOpen
  \bibfield  {author} {\bibinfo {author} {\bibfnamefont {J.}~\bibnamefont
  {Peñarrubia}}, \bibinfo {author} {\bibfnamefont {Y.-Z.}\ \bibnamefont {Ma}},
  \bibinfo {author} {\bibfnamefont {M.~G.}\ \bibnamefont {Walker}}, \ and\
  \bibinfo {author} {\bibfnamefont {A.}~\bibnamefont {McConnachie}},\ }\href
  {\doibase 10.1093/mnras/stu879} {\bibfield  {journal} {\bibinfo  {journal}
  {Mon. Not. Roy. Astron. Soc.}\ }\textbf {\bibinfo {volume} {443}},\ \bibinfo
  {pages} {2204} (\bibinfo {year} {2014})},\ \Eprint
  {http://arxiv.org/abs/1405.0306} {arXiv:1405.0306 [astro-ph.GA]} \BibitemShut
  {NoStop}%
\bibitem [{\citenamefont {Li}\ and\ \citenamefont {White}(2008)}]{Li:2007eg}%
  \BibitemOpen
  \bibfield  {author} {\bibinfo {author} {\bibfnamefont {Y.-S.}\ \bibnamefont
  {Li}}\ and\ \bibinfo {author} {\bibfnamefont {S.~D.~M.}\ \bibnamefont
  {White}},\ }\href {\doibase 10.1111/j.1365-2966.2007.12748.x} {\bibfield
  {journal} {\bibinfo  {journal} {Mon. Not. Roy. Astron. Soc.}\ }\textbf
  {\bibinfo {volume} {384}},\ \bibinfo {pages} {1459} (\bibinfo {year}
  {2008})},\ \Eprint {http://arxiv.org/abs/0710.3740} {arXiv:0710.3740
  [astro-ph]} \BibitemShut {NoStop}%
\bibitem [{\citenamefont {{Banik}}\ \emph {et~al.}(2018)\citenamefont
  {{Banik}}, \citenamefont {{O'Ryan}},\ and\ \citenamefont
  {{Zhao}}}]{2018MNRAS.477.4768B}%
  \BibitemOpen
  \bibfield  {author} {\bibinfo {author} {\bibfnamefont {I.}~\bibnamefont
  {{Banik}}}, \bibinfo {author} {\bibfnamefont {D.}~\bibnamefont {{O'Ryan}}}, \
  and\ \bibinfo {author} {\bibfnamefont {H.}~\bibnamefont {{Zhao}}},\ }\href
  {\doibase 10.1093/mnras/sty919} {\bibfield  {journal} {\bibinfo  {journal}
  {\mnras}\ }\textbf {\bibinfo {volume} {477}},\ \bibinfo {pages} {4768}
  (\bibinfo {year} {2018})},\ \Eprint {http://arxiv.org/abs/1802.00440}
  {arXiv:1802.00440 [astro-ph.GA]} \BibitemShut {NoStop}%
\bibitem [{\citenamefont {Hartl}\ and\ \citenamefont
  {Strigari}(2021)}]{Hartl:2021aio}%
  \BibitemOpen
  \bibfield  {author} {\bibinfo {author} {\bibfnamefont {O.~V.}\ \bibnamefont
  {Hartl}}\ and\ \bibinfo {author} {\bibfnamefont {L.~E.}\ \bibnamefont
  {Strigari}},\ }\href@noop {} {\  (\bibinfo {year} {2021})},\ \Eprint
  {http://arxiv.org/abs/2107.11490} {arXiv:2107.11490 [astro-ph.CO]}
  \BibitemShut {NoStop}%
\bibitem [{\citenamefont {van~der Marel}\ \emph {et~al.}(2012)\citenamefont
  {van~der Marel}, \citenamefont {Fardal}, \citenamefont {Besla}, \citenamefont
  {Beaton}, \citenamefont {Sohn}, \citenamefont {Anderson}, \citenamefont
  {Brown},\ and\ \citenamefont {Guhathakurta}}]{vanderMarel:2012xp}%
  \BibitemOpen
  \bibfield  {author} {\bibinfo {author} {\bibfnamefont {R.~P.}\ \bibnamefont
  {van~der Marel}}, \bibinfo {author} {\bibfnamefont {M.}~\bibnamefont
  {Fardal}}, \bibinfo {author} {\bibfnamefont {G.}~\bibnamefont {Besla}},
  \bibinfo {author} {\bibfnamefont {R.~L.}\ \bibnamefont {Beaton}}, \bibinfo
  {author} {\bibfnamefont {S.~T.}\ \bibnamefont {Sohn}}, \bibinfo {author}
  {\bibfnamefont {J.}~\bibnamefont {Anderson}}, \bibinfo {author}
  {\bibfnamefont {T.}~\bibnamefont {Brown}}, \ and\ \bibinfo {author}
  {\bibfnamefont {P.}~\bibnamefont {Guhathakurta}},\ }\href {\doibase
  10.1088/0004-637X/753/1/8} {\bibfield  {journal} {\bibinfo  {journal}
  {Astrophys. J.}\ }\textbf {\bibinfo {volume} {753}},\ \bibinfo {pages} {8}
  (\bibinfo {year} {2012})},\ \Eprint {http://arxiv.org/abs/1205.6864}
  {arXiv:1205.6864 [astro-ph.GA]} \BibitemShut {NoStop}%
\bibitem [{\citenamefont {Chernin}\ \emph {et~al.}(2009)\citenamefont
  {Chernin}, \citenamefont {Teerikorpi}, \citenamefont {Valtonen},
  \citenamefont {Byrd}, \citenamefont {Dolgachev},\ and\ \citenamefont
  {Domozhilova}}]{Chernin:2009ms}%
  \BibitemOpen
  \bibfield  {author} {\bibinfo {author} {\bibfnamefont {A.~D.}\ \bibnamefont
  {Chernin}}, \bibinfo {author} {\bibfnamefont {P.}~\bibnamefont {Teerikorpi}},
  \bibinfo {author} {\bibfnamefont {M.~J.}\ \bibnamefont {Valtonen}}, \bibinfo
  {author} {\bibfnamefont {G.~G.}\ \bibnamefont {Byrd}}, \bibinfo {author}
  {\bibfnamefont {V.~P.}\ \bibnamefont {Dolgachev}}, \ and\ \bibinfo {author}
  {\bibfnamefont {L.~M.}\ \bibnamefont {Domozhilova}},\ }\href@noop {} {\
  (\bibinfo {year} {2009})},\ \Eprint {http://arxiv.org/abs/0902.3871}
  {arXiv:0902.3871 [astro-ph.CO]} \BibitemShut {NoStop}%
\bibitem [{\citenamefont {Phelps}\ \emph {et~al.}(2013)\citenamefont {Phelps},
  \citenamefont {Nusser},\ and\ \citenamefont {Desjacques}}]{Phelps:2013rra}%
  \BibitemOpen
  \bibfield  {author} {\bibinfo {author} {\bibfnamefont {S.}~\bibnamefont
  {Phelps}}, \bibinfo {author} {\bibfnamefont {A.}~\bibnamefont {Nusser}}, \
  and\ \bibinfo {author} {\bibfnamefont {V.}~\bibnamefont {Desjacques}},\
  }\href {\doibase 10.1088/0004-637X/775/2/102} {\bibfield  {journal} {\bibinfo
   {journal} {Astrophys. J.}\ }\textbf {\bibinfo {volume} {775}},\ \bibinfo
  {pages} {102} (\bibinfo {year} {2013})},\ \Eprint
  {http://arxiv.org/abs/1306.4013} {arXiv:1306.4013 [astro-ph.CO]} \BibitemShut
  {NoStop}%
\bibitem [{\citenamefont {McLeod}\ \emph {et~al.}(2017)\citenamefont {McLeod},
  \citenamefont {Libeskind}, \citenamefont {Lahav},\ and\ \citenamefont
  {Hoffman}}]{McLeod:2016bjk}%
  \BibitemOpen
  \bibfield  {author} {\bibinfo {author} {\bibfnamefont {M.}~\bibnamefont
  {McLeod}}, \bibinfo {author} {\bibfnamefont {N.}~\bibnamefont {Libeskind}},
  \bibinfo {author} {\bibfnamefont {O.}~\bibnamefont {Lahav}}, \ and\ \bibinfo
  {author} {\bibfnamefont {Y.}~\bibnamefont {Hoffman}},\ }\href {\doibase
  10.1088/1475-7516/2017/12/034} {\bibfield  {journal} {\bibinfo  {journal}
  {JCAP}\ }\textbf {\bibinfo {volume} {1712}},\ \bibinfo {pages} {034}
  (\bibinfo {year} {2017})},\ \Eprint {http://arxiv.org/abs/1606.02694}
  {arXiv:1606.02694 [astro-ph.GA]} \BibitemShut {NoStop}%
\bibitem [{\citenamefont {Lemos}\ \emph {et~al.}(2021)\citenamefont {Lemos},
  \citenamefont {Jeffrey}, \citenamefont {Whiteway}, \citenamefont {Lahav},
  \citenamefont {Libeskind},\ and\ \citenamefont {Hoffman}}]{Lemos:2020vhj}%
  \BibitemOpen
  \bibfield  {author} {\bibinfo {author} {\bibfnamefont {P.}~\bibnamefont
  {Lemos}}, \bibinfo {author} {\bibfnamefont {N.}~\bibnamefont {Jeffrey}},
  \bibinfo {author} {\bibfnamefont {L.}~\bibnamefont {Whiteway}}, \bibinfo
  {author} {\bibfnamefont {O.}~\bibnamefont {Lahav}}, \bibinfo {author}
  {\bibfnamefont {N.~I.}\ \bibnamefont {Libeskind}}, \ and\ \bibinfo {author}
  {\bibfnamefont {Y.}~\bibnamefont {Hoffman}},\ }\href {\doibase
  10.1103/PhysRevD.103.023009} {\bibfield  {journal} {\bibinfo  {journal}
  {Phys. Rev. D}\ }\textbf {\bibinfo {volume} {103}},\ \bibinfo {pages}
  {023009} (\bibinfo {year} {2021})},\ \Eprint
  {http://arxiv.org/abs/2010.08537} {arXiv:2010.08537 [astro-ph.GA]}
  \BibitemShut {NoStop}%
\bibitem [{\citenamefont {Zhao}\ \emph {et~al.}(2013)\citenamefont {Zhao},
  \citenamefont {Famaey}, \citenamefont {Lüghausen},\ and\ \citenamefont
  {Kroupa}}]{Zhao:2013uya}%
  \BibitemOpen
  \bibfield  {author} {\bibinfo {author} {\bibfnamefont {H.}~\bibnamefont
  {Zhao}}, \bibinfo {author} {\bibfnamefont {B.}~\bibnamefont {Famaey}},
  \bibinfo {author} {\bibfnamefont {F.}~\bibnamefont {Lüghausen}}, \ and\
  \bibinfo {author} {\bibfnamefont {P.}~\bibnamefont {Kroupa}},\ }\href
  {\doibase 10.1051/0004-6361/201321879} {\bibfield  {journal} {\bibinfo
  {journal} {Astron. Astrophys.}\ }\textbf {\bibinfo {volume} {557}},\ \bibinfo
  {pages} {L3} (\bibinfo {year} {2013})},\ \Eprint
  {http://arxiv.org/abs/1306.6628} {arXiv:1306.6628 [astro-ph.GA]} \BibitemShut
  {NoStop}%
\bibitem [{\citenamefont {Benisty}\ and\ \citenamefont
  {Guendelman}(2020)}]{Benisty:2020kys}%
  \BibitemOpen
  \bibfield  {author} {\bibinfo {author} {\bibfnamefont {D.}~\bibnamefont
  {Benisty}}\ and\ \bibinfo {author} {\bibfnamefont {E.~I.}\ \bibnamefont
  {Guendelman}},\ }\href {\doibase 10.1016/j.dark.2020.100708} {\bibfield
  {journal} {\bibinfo  {journal} {Phys. Dark Univ.}\ }\textbf {\bibinfo
  {volume} {30}},\ \bibinfo {pages} {100708} (\bibinfo {year} {2020})},\
  \Eprint {http://arxiv.org/abs/2007.13006} {arXiv:2007.13006 [astro-ph.CO]}
  \BibitemShut {NoStop}%
\bibitem [{\citenamefont {Chernin}\ \emph {et~al.}(2000)\citenamefont
  {Chernin}, \citenamefont {Teerikorpi},\ and\ \citenamefont
  {Baryshev}}]{Chernin:2000pq}%
  \BibitemOpen
  \bibfield  {author} {\bibinfo {author} {\bibfnamefont {A.}~\bibnamefont
  {Chernin}}, \bibinfo {author} {\bibfnamefont {P.}~\bibnamefont {Teerikorpi}},
  \ and\ \bibinfo {author} {\bibfnamefont {Y.}~\bibnamefont {Baryshev}},\
  }\href@noop {} {\  (\bibinfo {year} {2000})},\ \Eprint
  {http://arxiv.org/abs/astro-ph/0012021} {arXiv:astro-ph/0012021} \BibitemShut
  {NoStop}%
\bibitem [{\citenamefont {Baryshev}\ \emph {et~al.}(2000)\citenamefont
  {Baryshev}, \citenamefont {Chernin},\ and\ \citenamefont
  {Teerikorpi}}]{Baryshev:2000kw}%
  \BibitemOpen
  \bibfield  {author} {\bibinfo {author} {\bibfnamefont {Y.}~\bibnamefont
  {Baryshev}}, \bibinfo {author} {\bibfnamefont {A.}~\bibnamefont {Chernin}}, \
  and\ \bibinfo {author} {\bibfnamefont {P.}~\bibnamefont {Teerikorpi}},\
  }\href@noop {} {\  (\bibinfo {year} {2000})},\ \Eprint
  {http://arxiv.org/abs/astro-ph/0011528} {arXiv:astro-ph/0011528} \BibitemShut
  {NoStop}%
\bibitem [{\citenamefont {Chernin}\ \emph {et~al.}(2002)\citenamefont
  {Chernin}, \citenamefont {Santiago},\ and\ \citenamefont
  {Silbergleit}}]{Chernin:2001nu}%
  \BibitemOpen
  \bibfield  {author} {\bibinfo {author} {\bibfnamefont {A.~D.}\ \bibnamefont
  {Chernin}}, \bibinfo {author} {\bibfnamefont {D.~I.}\ \bibnamefont
  {Santiago}}, \ and\ \bibinfo {author} {\bibfnamefont {A.~S.}\ \bibnamefont
  {Silbergleit}},\ }\href {\doibase 10.1016/S0375-9601(01)00672-7} {\bibfield
  {journal} {\bibinfo  {journal} {Phys. Lett. A}\ }\textbf {\bibinfo {volume}
  {294}},\ \bibinfo {pages} {79} (\bibinfo {year} {2002})},\ \Eprint
  {http://arxiv.org/abs/astro-ph/0106144} {arXiv:astro-ph/0106144} \BibitemShut
  {NoStop}%
\bibitem [{\citenamefont {Karachentsev}\ \emph {et~al.}(2003)\citenamefont
  {Karachentsev}, \citenamefont {Chernin},\ and\ \citenamefont
  {Teerikorpi}}]{Karachentsev:2003eh}%
  \BibitemOpen
  \bibfield  {author} {\bibinfo {author} {\bibfnamefont {I.~D.}\ \bibnamefont
  {Karachentsev}}, \bibinfo {author} {\bibfnamefont {A.~D.}\ \bibnamefont
  {Chernin}}, \ and\ \bibinfo {author} {\bibfnamefont {P.}~\bibnamefont
  {Teerikorpi}},\ }\href@noop {} {\bibfield  {journal} {\bibinfo  {journal}
  {Astrofiz.}\ }\textbf {\bibinfo {volume} {46}},\ \bibinfo {pages} {399}
  (\bibinfo {year} {2003})},\ \Eprint {http://arxiv.org/abs/astro-ph/0304250}
  {arXiv:astro-ph/0304250} \BibitemShut {NoStop}%
\bibitem [{\citenamefont {Chernin}\ \emph {et~al.}(2004)\citenamefont
  {Chernin}, \citenamefont {Karachentsev}, \citenamefont {Valtonen},
  \citenamefont {Dolgachev}, \citenamefont {Domozhilova},\ and\ \citenamefont
  {Makarov}}]{Chernin:2003qd}%
  \BibitemOpen
  \bibfield  {author} {\bibinfo {author} {\bibfnamefont {A.~D.}\ \bibnamefont
  {Chernin}}, \bibinfo {author} {\bibfnamefont {I.~D.}\ \bibnamefont
  {Karachentsev}}, \bibinfo {author} {\bibfnamefont {M.~J.}\ \bibnamefont
  {Valtonen}}, \bibinfo {author} {\bibfnamefont {V.~P.}\ \bibnamefont
  {Dolgachev}}, \bibinfo {author} {\bibfnamefont {L.~M.}\ \bibnamefont
  {Domozhilova}}, \ and\ \bibinfo {author} {\bibfnamefont {D.~I.}\ \bibnamefont
  {Makarov}},\ }\href {\doibase 10.1051/0004-6361:20034170} {\bibfield
  {journal} {\bibinfo  {journal} {Astron. Astrophys.}\ }\textbf {\bibinfo
  {volume} {415}},\ \bibinfo {pages} {19} (\bibinfo {year} {2004})},\ \Eprint
  {http://arxiv.org/abs/astro-ph/0310048} {arXiv:astro-ph/0310048} \BibitemShut
  {NoStop}%
\bibitem [{\citenamefont {Teerikorpi}\ \emph {et~al.}(2005)\citenamefont
  {Teerikorpi}, \citenamefont {Chernin},\ and\ \citenamefont
  {Baryshev}}]{Teerikorpi:2005zh}%
  \BibitemOpen
  \bibfield  {author} {\bibinfo {author} {\bibfnamefont {P.}~\bibnamefont
  {Teerikorpi}}, \bibinfo {author} {\bibfnamefont {A.~D.}\ \bibnamefont
  {Chernin}}, \ and\ \bibinfo {author} {\bibfnamefont {Y.~V.}\ \bibnamefont
  {Baryshev}},\ }\href {\doibase 10.1051/0004-6361:20053139} {\bibfield
  {journal} {\bibinfo  {journal} {Astron. Astrophys.}\ }\textbf {\bibinfo
  {volume} {440}},\ \bibinfo {pages} {791} (\bibinfo {year} {2005})},\ \Eprint
  {http://arxiv.org/abs/astro-ph/0506683} {arXiv:astro-ph/0506683} \BibitemShut
  {NoStop}%
\bibitem [{\citenamefont {Chernin}\ \emph {et~al.}(2006)\citenamefont
  {Chernin}, \citenamefont {Teerikorpi},\ and\ \citenamefont
  {Baryshev}}]{Chernin:2006dy}%
  \BibitemOpen
  \bibfield  {author} {\bibinfo {author} {\bibfnamefont {A.~D.}\ \bibnamefont
  {Chernin}}, \bibinfo {author} {\bibfnamefont {P.}~\bibnamefont {Teerikorpi}},
  \ and\ \bibinfo {author} {\bibfnamefont {Y.~V.}\ \bibnamefont {Baryshev}},\
  }\href {\doibase 10.1051/0004-6361:20054668} {\bibfield  {journal} {\bibinfo
  {journal} {Astron. Astrophys.}\ }\textbf {\bibinfo {volume} {456}},\ \bibinfo
  {pages} {13} (\bibinfo {year} {2006})},\ \Eprint
  {http://arxiv.org/abs/astro-ph/0603226} {arXiv:astro-ph/0603226} \BibitemShut
  {NoStop}%
\bibitem [{\citenamefont {Teerikorpi}\ and\ \citenamefont
  {Chernin}(2010)}]{Teerikorpi:2010zz}%
  \BibitemOpen
  \bibfield  {author} {\bibinfo {author} {\bibfnamefont {P.}~\bibnamefont
  {Teerikorpi}}\ and\ \bibinfo {author} {\bibfnamefont {A.~D.}\ \bibnamefont
  {Chernin}},\ }\href {\doibase 10.1051/0004-6361/201014346} {\bibfield
  {journal} {\bibinfo  {journal} {Astron. Astrophys.}\ }\textbf {\bibinfo
  {volume} {516}},\ \bibinfo {pages} {A93} (\bibinfo {year} {2010})},\ \Eprint
  {http://arxiv.org/abs/1006.0066} {arXiv:1006.0066 [astro-ph.CO]} \BibitemShut
  {NoStop}%
\bibitem [{\citenamefont {Chernin}(2015)}]{Chernin:2015nna}%
  \BibitemOpen
  \bibfield  {author} {\bibinfo {author} {\bibfnamefont {A.~D.}\ \bibnamefont
  {Chernin}},\ }\bibfield  {booktitle} {\emph {\bibinfo {booktitle}
  {{Proceedings, Subatomic particles, Nucleons, Atoms, Universe: Processes and
  Structure: International conference in honor of Ya. B. Zeldovich 100th
  Anniversary: Minsk, Belarus, March 10-14, 2014}}},\ }\href {\doibase
  10.1134/S1063772915060104} {\bibfield  {journal} {\bibinfo  {journal}
  {Astron. Rep.}\ }\textbf {\bibinfo {volume} {59}},\ \bibinfo {pages} {474}
  (\bibinfo {year} {2015})}\BibitemShut {NoStop}%
\bibitem [{\citenamefont {Silbergleit}\ and\ \citenamefont
  {Chernin}(2019)}]{Silbergleit:2019oyx}%
  \BibitemOpen
  \bibfield  {author} {\bibinfo {author} {\bibfnamefont {A.}~\bibnamefont
  {Silbergleit}}\ and\ \bibinfo {author} {\bibfnamefont {A.}~\bibnamefont
  {Chernin}},\ }\href {\doibase 10.1007/978-3-030-36752-7} {\emph {\bibinfo
  {title} {{Kepler Problem in the Presence of Dark Energy, and the Cosmic Local
  Flow}}}},\ SpringerBriefs in Physics\ (\bibinfo  {publisher} {Springer},\
  \bibinfo {year} {2019})\BibitemShut {NoStop}%
\bibitem [{\citenamefont {Emelyanov}\ \emph {et~al.}(2015)\citenamefont
  {Emelyanov}, \citenamefont {Kovalyov},\ and\ \citenamefont
  {Chernin}}]{Emelyanov:2015ina}%
  \BibitemOpen
  \bibfield  {author} {\bibinfo {author} {\bibfnamefont {N.~V.}\ \bibnamefont
  {Emelyanov}}, \bibinfo {author} {\bibfnamefont {M.~{\relax Yu}.}\
  \bibnamefont {Kovalyov}}, \ and\ \bibinfo {author} {\bibfnamefont {A.~D.}\
  \bibnamefont {Chernin}},\ }\href {\doibase 10.1134/S1063772915050029}
  {\bibfield  {journal} {\bibinfo  {journal} {Astron. Rep.}\ }\textbf {\bibinfo
  {volume} {59}},\ \bibinfo {pages} {510} (\bibinfo {year} {2015})}\BibitemShut
  {NoStop}%
\bibitem [{\citenamefont {Carrera}\ and\ \citenamefont
  {Giulini}(2006)}]{Carrera:2006im}%
  \BibitemOpen
  \bibfield  {author} {\bibinfo {author} {\bibfnamefont {M.}~\bibnamefont
  {Carrera}}\ and\ \bibinfo {author} {\bibfnamefont {D.}~\bibnamefont
  {Giulini}},\ }\href@noop {} {\  (\bibinfo {year} {2006})},\ \Eprint
  {http://arxiv.org/abs/gr-qc/0602098} {arXiv:gr-qc/0602098 [gr-qc]}
  \BibitemShut {NoStop}%
\bibitem [{\citenamefont {Aghanim}\ \emph {et~al.}(2020)\citenamefont {Aghanim}
  \emph {et~al.}}]{Aghanim:2018eyx}%
  \BibitemOpen
  \bibfield  {author} {\bibinfo {author} {\bibfnamefont {N.}~\bibnamefont
  {Aghanim}} \emph {et~al.} (\bibinfo {collaboration} {Planck}),\ }\href
  {\doibase 10.1051/0004-6361/201833910} {\bibfield  {journal} {\bibinfo
  {journal} {Astron. Astrophys.}\ }\textbf {\bibinfo {volume} {641}},\ \bibinfo
  {pages} {A6} (\bibinfo {year} {2020})},\ \bibinfo {note} {[Erratum:
  Astron.Astrophys. 652, C4 (2021)]},\ \Eprint
  {http://arxiv.org/abs/1807.06209} {arXiv:1807.06209 [astro-ph.CO]}
  \BibitemShut {NoStop}%
\bibitem [{\citenamefont {Eingorn}\ \emph {et~al.}(2013)\citenamefont
  {Eingorn}, \citenamefont {Kudinova},\ and\ \citenamefont
  {Zhuk}}]{Eingorn:2012dg}%
  \BibitemOpen
  \bibfield  {author} {\bibinfo {author} {\bibfnamefont {M.}~\bibnamefont
  {Eingorn}}, \bibinfo {author} {\bibfnamefont {A.}~\bibnamefont {Kudinova}}, \
  and\ \bibinfo {author} {\bibfnamefont {A.}~\bibnamefont {Zhuk}},\ }\href
  {\doibase 10.1088/1475-7516/2013/04/010} {\bibfield  {journal} {\bibinfo
  {journal} {JCAP}\ }\textbf {\bibinfo {volume} {1304}},\ \bibinfo {pages}
  {010} (\bibinfo {year} {2013})},\ \Eprint {http://arxiv.org/abs/1211.4045}
  {arXiv:1211.4045 [astro-ph.CO]} \BibitemShut {NoStop}%
\bibitem [{\citenamefont {Eingorn}\ and\ \citenamefont
  {Zhuk}(2012)}]{Eingorn:2012jm}%
  \BibitemOpen
  \bibfield  {author} {\bibinfo {author} {\bibfnamefont {M.}~\bibnamefont
  {Eingorn}}\ and\ \bibinfo {author} {\bibfnamefont {A.}~\bibnamefont {Zhuk}},\
  }\href {\doibase 10.1088/1475-7516/2012/09/026} {\bibfield  {journal}
  {\bibinfo  {journal} {JCAP}\ }\textbf {\bibinfo {volume} {09}},\ \bibinfo
  {pages} {026} (\bibinfo {year} {2012})},\ \Eprint
  {http://arxiv.org/abs/1205.2384} {arXiv:1205.2384 [astro-ph.CO]} \BibitemShut
  {NoStop}%
\bibitem [{\citenamefont {Partridge}\ \emph {et~al.}(2013)\citenamefont
  {Partridge}, \citenamefont {Lahav},\ and\ \citenamefont
  {Hoffman}}]{Partridge:2013dsa}%
  \BibitemOpen
  \bibfield  {author} {\bibinfo {author} {\bibfnamefont {C.}~\bibnamefont
  {Partridge}}, \bibinfo {author} {\bibfnamefont {O.}~\bibnamefont {Lahav}}, \
  and\ \bibinfo {author} {\bibfnamefont {Y.}~\bibnamefont {Hoffman}},\ }\href
  {\doibase 10.1093/mnrasl/slt109} {\bibfield  {journal} {\bibinfo  {journal}
  {Mon. Not. Roy. Astron. Soc.}\ }\textbf {\bibinfo {volume} {436}},\ \bibinfo
  {pages} {45} (\bibinfo {year} {2013})},\ \Eprint
  {http://arxiv.org/abs/1308.0970} {arXiv:1308.0970 [astro-ph.CO]} \BibitemShut
  {NoStop}%
\bibitem [{\citenamefont {McLeod}\ and\ \citenamefont
  {Lahav}(2020)}]{McLeod:2019cfg}%
  \BibitemOpen
  \bibfield  {author} {\bibinfo {author} {\bibfnamefont {M.}~\bibnamefont
  {McLeod}}\ and\ \bibinfo {author} {\bibfnamefont {O.}~\bibnamefont {Lahav}},\
  }\href {\doibase 10.1088/1475-7516/2020/09/056} {\bibfield  {journal}
  {\bibinfo  {journal} {JCAP}\ }\textbf {\bibinfo {volume} {09}},\ \bibinfo
  {pages} {056} (\bibinfo {year} {2020})},\ \Eprint
  {http://arxiv.org/abs/1903.10849} {arXiv:1903.10849 [astro-ph.CO]}
  \BibitemShut {NoStop}%
\bibitem [{\citenamefont {{Binney}}\ and\ \citenamefont
  {{Tremaine}}(1987)}]{1987gady}%
  \BibitemOpen
  \bibfield  {author} {\bibinfo {author} {\bibfnamefont {J.}~\bibnamefont
  {{Binney}}}\ and\ \bibinfo {author} {\bibfnamefont {S.}~\bibnamefont
  {{Tremaine}}},\ }\href@noop {} {\emph {\bibinfo {title} {Princeton, NJ,
  Princeton University Press, 1987, 747 p.}}}\ (\bibinfo {year}
  {1987})\BibitemShut {NoStop}%
\bibitem [{\citenamefont {Hammer}\ \emph {et~al.}(2007)\citenamefont {Hammer},
  \citenamefont {Puech}, \citenamefont {Chemin}, \citenamefont {Flores},\ and\
  \citenamefont {Lehnert}}]{Hammer:2007ki}%
  \BibitemOpen
  \bibfield  {author} {\bibinfo {author} {\bibfnamefont {F.}~\bibnamefont
  {Hammer}}, \bibinfo {author} {\bibfnamefont {M.}~\bibnamefont {Puech}},
  \bibinfo {author} {\bibfnamefont {L.}~\bibnamefont {Chemin}}, \bibinfo
  {author} {\bibfnamefont {H.}~\bibnamefont {Flores}}, \ and\ \bibinfo {author}
  {\bibfnamefont {M.}~\bibnamefont {Lehnert}},\ }\href {\doibase
  10.1086/516727} {\bibfield  {journal} {\bibinfo  {journal} {Astrophys. J.}\
  }\textbf {\bibinfo {volume} {662}},\ \bibinfo {pages} {322} (\bibinfo {year}
  {2007})},\ \Eprint {http://arxiv.org/abs/astro-ph/0702585}
  {arXiv:astro-ph/0702585 [astro-ph]} \BibitemShut {NoStop}%
\bibitem [{\citenamefont {Cox}\ and\ \citenamefont {Loeb}(2008)}]{Cox:2007nt}%
  \BibitemOpen
  \bibfield  {author} {\bibinfo {author} {\bibfnamefont {T.~J.}\ \bibnamefont
  {Cox}}\ and\ \bibinfo {author} {\bibfnamefont {A.}~\bibnamefont {Loeb}},\
  }\href {\doibase 10.1111/j.1365-2966.2008.13048.x} {\bibfield  {journal}
  {\bibinfo  {journal} {Mon. Not. Roy. Astron. Soc.}\ }\textbf {\bibinfo
  {volume} {386}},\ \bibinfo {pages} {461} (\bibinfo {year} {2008})},\ \Eprint
  {http://arxiv.org/abs/0705.1170} {arXiv:0705.1170 [astro-ph]} \BibitemShut
  {NoStop}%
\bibitem [{\citenamefont {Conselice}\ \emph {et~al.}(2009)\citenamefont
  {Conselice}, \citenamefont {Yang},\ and\ \citenamefont {Bluck}}]{Conselice}%
  \BibitemOpen
  \bibfield  {author} {\bibinfo {author} {\bibfnamefont {C.~J.}\ \bibnamefont
  {Conselice}}, \bibinfo {author} {\bibfnamefont {C.}~\bibnamefont {Yang}}, \
  and\ \bibinfo {author} {\bibfnamefont {A.~F.~L.}\ \bibnamefont {Bluck}},\
  }\href {\doibase 10.1111/j.1365-2966.2009.14396.x} {\bibfield  {journal}
  {\bibinfo  {journal} {Monthly Notices of the Royal Astronomical Society}\
  }\textbf {\bibinfo {volume} {394}},\ \bibinfo {pages} {1956} (\bibinfo {year}
  {2009})}\BibitemShut {NoStop}%
\end{thebibliography}%
\bibliographystyle{apsrev4-1}

\end{document}